\documentclass[10pt,conference]{IEEEtran}

\IEEEoverridecommandlockouts
\usepackage{cite}
\usepackage{amsmath,amssymb,amsfonts}
\usepackage{algorithmic}
\usepackage{graphicx}
\usepackage{textcomp}
\usepackage{booktabs}
\usepackage{multirow}
\usepackage{xcolor}
\usepackage{url}
\usepackage{listings}
\usepackage{tcolorbox}
\usepackage{balance}
\newtcolorbox{rqanswer}[1]{colback=gray!7, colframe=gray!55, boxrule=0.4pt,
  arc=1pt, left=4pt, right=4pt, top=3pt, bottom=3pt, before skip=6pt,
  after skip=6pt, fontupper=\small, before upper={\textbf{Answer to #1.}~}}
\def\BibTeX{{\rm B\kern-.05em{\sc i\kern-.025em b}\kern-.08em
    T\kern-.1667em\lower.7ex\hbox{E}\kern-.125emX}}

\begin{document}


\title{Which Signals of Code Correctness Survive Distribution Shift? Hidden-State and Causal Predictors of Program and Verdict Correctness}



\author{\IEEEauthorblockN{
Viola Campos, Thomas Klassert, and Adrian Ulges
}
\IEEEauthorblockA{\textit{RheinMain University of Applied Sciences}\\
Wiesbaden, Germany \\
\{viola.campos, adrian.ulges\}@hs-rm.de} thomas.klassert@student.hs-rm.de
}

\maketitle

\begin{abstract}
Large language models increasingly judge code, 
yet without external feedback such judgments are unreliable, and verifying them by
executing tests or by extended reasoning is costly. 
Instead, execution-free correctness predictors use signals from the model's output probabilities, sampled solutions, or hidden states to identify judgments that warrant further verification.
However, 
these predictors 
may degrade under distribution
shifts 
when generalizing across projects and languages. 

We compare six execution-free predictors from four families on Qwen3-Coder-30B and gpt-oss-20b under benchmark shift between LiveCodeBench and BigCodeBench and language shift across Python, C++ and Java.
We evaluate two targets: whether the model's own verdict on a program can be trusted, and whether the program itself is correct.
Besides established approaches, we adapt two novel causal localization methods to code, which have been shown to be more robust out-of-distribution on natural language tasks. 

For verdict correctness, causal localization improves robustness to language shift. For Qwen3-Coder a probe on the causally localized subspace stays 0.09 AUROC ahead of output confidence under language shift (0.11 in distribution), significantly outperforming the best non-causal predictor by 0.04 AUROC on average.
This advantage, however, does not generalize consistently to gpt-oss or to benchmark shift. For program correctness, predictors transfer more successfully between benchmarks, largely by identifying problems on which the model tends to fail. However, their ability to rank candidates within the same problem is limited. 

\end{abstract}

\begin{IEEEkeywords}
code correctness prediction, uncertainty estimation, LLM-as-a-judge, distribution shift, causal interpretability
\end{IEEEkeywords}

\section{Introduction}
\label{sec:intro}
LLMs are increasingly used to judge code, including their own. They determine whether a candidate solves
its task, whether a change is safe to accept, and which of several candidates to
keep. Without external feedback, such judgments are unreliable. Models rarely
improve by verifying their own outputs~\cite{Huang2024SelfCorrect,
Kamoi2024SelfCorrection}, favor their own generations~\cite{Panickssery2024Self},
and remain inconsistent as code judges~\cite{Jiang2025CodeJudgeBench}. Reliable
verification is costly: executing tests requires an environment and a test
suite~\cite{Raghavendra2026Rubrics}, and reasoning about a candidate involves generating many additional
tokens. A cheap signal indicating when a judgment can be trusted could gate
these costlier checks rather than replace them~\cite{Rakasi2026Uncertainty,
jung2025trust}. Execution-free correctness predictors provide such a signal from
the model itself, such as the probability of its
answer~\cite{Spiess2025Calib, Campos2026Multicalib}, the agreement among sampled
solutions~\cite{Chen24Eigenscore, shi2026codemorethantext,
bouchard2026functional}, or its hidden states~\cite{bui2025openia,
Vu2026Autoprobe, ribeiro2026repe}. Probes on hidden states are particularly
cheap, since they reuse activations the model computes anyway, and are already
deployed as first-stage monitors~\cite{McKenzie2025HighStakes,
Kramar2026Probes}. Such a predictor serves two purposes: it can estimate whether
a program is correct, and whether the model's own verdict on it can be trusted.

Although predictors trained on hidden states are accurate in distribution, their
acknowledged weakness is distribution shift~\cite{Kramar2026Probes,
OrgadTGRSKB25}. For code this is critical, as a predictor fitted on one benchmark
or programming language will usually be applied to other projects and languages. For natural-language question answering, Huang et
al.~\cite{Huang2025Internal} proposed causal localization as a remedy: they use
counterfactual interventions, which copy part of the hidden state from one
input into another and observe how the model's answer changes, to find the
subspace that robustly determines the answer 
under shift. Whether this promise holds for code correctness assessment 
remains open. Instead, prior work on code has considered only a subset of predictor families
and shift settings, and has largely focused on predicting program
correctness~\cite{ribeiro2026robustness, he2026codecircuit, huang2025pttrust,
gros2025localized}. To our knowledge, none includes causal methods or asks whether the model's
own verdict can be trusted.

We therefore ask \emph{which execution-free signals of code correctness
survive a change of benchmark or programming language, and what explains those
that do}. We adapt two causal methods of Huang et al., \emph{counterfactual
simulation} and \emph{value probing}, to code and compare them with output
confidence, a probe on the hidden state, LAT~\cite{ribeiro2026repe} and
EigenScore~\cite{Chen24Eigenscore}. To do so, we use Qwen3-Coder-30B and gpt-oss-20b for candidate generation and as judges; we apply shifts between LiveCodeBench and BigCodeBench
and between Python, C++ and Java. We then answer four research questions: \textbf{RQ1} asks how well the methods tell
whether the model's verdict can be trusted, the target the causal methods were
designed for; \textbf{RQ2} asks which part of the causal methods carries their
signal; \textbf{RQ3} asks how well the methods predict whether the program is
correct, the target of most prior work; and \textbf{RQ4} asks whether
agreement across samples, which requires no supervised training, tells better than a single
pass which problems the model will solve. 

Our results show that causal localization pays off where it was designed to:
under language shift, a probe on the localized subspace is the most robust
predictor of Qwen3-Coder's verdict, significantly ahead of every non-causal
method, and the learned subspace itself, not the intervention, carries this
gain. On gpt-oss, no method stands out. Beyond the causal methods,
generalization depends on the target and the type of shift: no verdict
predictor transfers reliably between benchmarks, program correctness
transfers largely because the predictors recognize difficult problems, and
agreement across samples gives no consistent advantage over a single pass.


Our contributions are:
\begin{itemize}
  \item To our knowledge, the first comparison of execution-free correctness
  predictors for code across four families, including causal methods, under
  both benchmark and language shift and for two targets.
  \item An adaptation of causal localization~\cite{Huang2025Internal} to code
  correctness prediction.
  \item An analysis showing that the learned subspace carries the advantage of
  the causal methods, and a distinction between candidate-level prediction and
  problem-level reliability: pooled AUROC must be read alongside a predictor's
  ability to rank the candidates of one problem.
\end{itemize}

\section{Related Work}
\label{sec:related}

\paragraph{Code correctness prediction}
Token probabilities form the most direct signal for execution-free code correctness prediction. They have been studied for calibration~\cite{Spiess2025Calib,Campos2026Multicalib}, but have been found to barely separate failing from passing programs~\cite{rathnasuriya2026overconfident}. Consistency methods compare multiple  generations for the same problem, some by executing them~\cite{valentin2026incoherence,sharma2025,he2026usingsemanticdistanceestimate}, others without execution~\cite{jain2023lightweight,shi2026codemorethantext,bouchard2026functional}. Probes and representation reading recover correctness from hidden states~\cite{bui2025openia,Vu2026Autoprobe,ribeiro2026repe,huang2025pttrust,lugoloobi2026llms}.
Comparisons across the above approaches only exist between output and consistency methods so far~\cite{bouchard2026functional}. 
To our knowledge, no causal methods have been investigated.

\paragraph{Verdict correctness prediction}
LLMs are increasingly used as judges in software engineering~\cite{he2025judgese}, but
 have been found to misjudge code correctness~\cite{crupi2025judge} and react to superficial edits such as renamed variables~\cite{moon2026cover}. 
Hence, {\it verdict correctness prediction} aims at predicting the correctness not of the code itself, but of the LLM's asssessment. 
Approaches have explored 
judges' hidden states~\cite{Radharapu2025CalibratingJudges}, reasoning structure~\cite{qi2026verdi}, or have escalated uncertain cases~\cite{jung2025trust}. 
Furthermore, probes trained for correctness can track the model's own judgment instead~\cite{lu2026diagnosing}, which supports keeping the verdict and program correctness as separate targets. To our knowledge, however, no prior work studies both code correctness and verdict correctness, nor has verdict correctness been assessed under distribution shift. 



\paragraph{Generalization under shift}
ML models in software engineering are well known to erode for cross-dataset and cross-project prediction~\cite{zimmermann2009cpdp,hosseini2019cpdp}, under language shift~\cite{he2026codecircuit,huang2025pttrust},
and under other realistic splits~\cite{ding2025primevul}.
This has also been found for recent LLM-based predictors ~\cite{OrgadTGRSKB25,Kramar2026Probes,Srey2026Factorised}, particularly for 
execution-free code correctness prediction:  Probes based on hidden states have been found to lack robustness~\cite{ribeiro2026robustness}, and to fall behind under dataset shift~\cite{gros2025localized}.
Each of these works studies one or two families of predictors on program correctness; we compare four families of predictors, evaluate not only code correctness but also verdict correctness, and include both benchmark and language shifts.

\paragraph{Causal localization}
Huang et al.~\cite{Huang2025Internal} use causal localization 
with distributed alignment search~\cite{GeigerWPIG24DAS} to learn a subspace through which interchange interventions~\cite{Geiger2021CausalAbstractions} control the output, and use it for correctness prediction,  reporting strong robustness under shift on natural-language tasks. 
In general, whether a causal role makes a direction generalize better is open: some work finds that it does~\cite{marks2024geometry,mallen2024quirky}, while others find that causal efficacy and generality diverge~\cite{ying2026truthfulness}.
To this end, our work is the first to adapt and evaluate causal probing for code correctness and verdict estimation.

That an intervention changes the output does not by itself show that the model represents the variable there: patched subspaces can act through pathways the model does not normally use, and they need not be unique~\cite{makelov2024illusion,grant2025divergent,meloux2025everything}. We therefore compare the learned subspace with random subspaces of the same dimension 
(Section~\ref{sec:rq-subspace}).


\section{Problem Setting}
\label{sec:problem}
We study execution-free correctness prediction for code that an LLM generates
and judges itself, from signals of the model's own generation process: no tests are
executed and no additional reasoning tokens are generated. Given a programming
problem $x_i$, a model $\mathcal{M}$ samples candidate solutions
$c_{ij} \sim \mathcal{M}(x_i)$, $j = 1, \dots, n_i$. Following common practice in code generation benchmarks, a candidate is
considered \emph{correct} if it passes all tests. We denote the program correctness label as $y^{\mathrm{prog}}_{ij} \in \{0,1\}$.
Tests are executed only to obtain these labels for training and evaluation,
never to make a prediction. Where a statement holds for any problem and
candidate, we write $(x, c)$ and drop the indices.

\subsection{The Model's Verdict} 
\label{sec:problem-verdict}
To obtain the model's own judgment, we prompt $\mathcal{M}$ with the problem description $x_i$ and the candidate solution $c_{ij}$
and ask whether the implementation is correct, explicitely demanding a single token answer, \texttt{Yes} or \texttt{No}. We define the model's \emph{verdict} $v(x, c)$ as the
answer $t \in \{\texttt{Yes}, \texttt{No}\}$ with the higher probability
$p_{\mathcal{M}}(t {\mid} x, c)$. We refer to this probability as the {\it verdict confidence} and use it as a baseline in our experiments. The same forward pass that defines the verdict -- the \emph{judging
pass} --
 also generates the signals for the other prediction methods (Section~\ref{sec:methods}). We extract those at the verdict token's position: 
probabilities of the two answer tokens {\tt Yes/No}, the hidden states, or a subspace of
the hidden states identified through interventions. 

\subsection{Prediction Targets}
We evaluate predictors for two different targets: 

\paragraph{Program correctness ($y^{\mathrm{prog}}$)}
The first target is to predict whether the candidate implementation actually satisfies the problem specification, as operationalized by the test suite. This target corresponds to prior work on code correctness
prediction~\cite{bui2025openia, ribeiro2026repe, valentin2026incoherence,
sharma2025, he2026usingsemanticdistanceestimate}, and 
prediction scores are used to decide which candidates need testing.
Note that predictors may differ from the model's verdict $v$ as defined in Section \ref{sec:problem-verdict}, since they can be based on other internal signals. 

\paragraph{Verdict correctness ($y^{\mathrm{verd}}$)}
The second target is whether the model's verdict (as defined in Section \ref{sec:problem-verdict}) is correct, that is, whether
the model accepts a correct candidate or rejects an incorrect one:
\begin{equation*}
y^{\mathrm{verd}} = \mathbf{1}\bigl[v(x, c) = \texttt{Yes} \Leftrightarrow
y^{\mathrm{prog}} = 1\bigr].
\end{equation*}
A predictor for this target estimates when the model's judgment can be
trusted and when to invoke more expensive verification, such as test
execution or a stronger model. It is related to uncertainty estimation for generated answers, which predicts whether a model's own answer is
correct~\cite{Kadavath2022LanguageM, Huang2025Internal}.

Note that these two targets serve different purposes: program correctness concerns the
candidate itself, whereas verdict correctness concerns the reliability of the model's
judgment. When a model correctly rejects an incorrect candidate, for example,
its verdict is correct while the program is not. We evaluate both targets with the
same judging pass to investigate which internal signals support each prediction target.

\subsection{Distribution Shift}
\label{sec:problem-shift}
We study two
generalization scenarios, in which the evaluation data differ from the
training data in one controlled way.

In \emph{benchmark shift}, we train a predictor on one benchmark,
LiveCodeBench~\cite{Jain2025Livecodebench} or
BigCodeBench~\cite{Zhuo2025BCB}, and evaluate it on the other. This simulates
a predictor fitted on one kind of task and deployed on another:
LiveCodeBench consists of algorithmic programming challenges, BigCodeBench of
practical tasks that call libraries and APIs.

In \emph{language shift}, we train a predictor on candidates in one
programming language and evaluate it on candidates in another, for the same
problems in Python, C++ and Java. The problems are split once for all three
languages, so no problem is seen in training in one language and in
evaluation in another. 

\section{Methods}
\label{sec:methods}
We compare six predictors from four families, grouped by the signal they read
from the model: (1) its output token probabilities, (2) the consistency across
sampled generations, (3) internal features of the hidden states, and (4)
internal causal features, a subspace of the hidden states identified through counterfactual interventions. The first three families are established approaches to
correctness prediction. The fourth was proposed for natural-language question
answering by Huang et al.~\cite{Huang2025Internal}, who report that it is more robust under distribution shift. We adapt this approach to code generation 
and describe it in more detail.

\subsection{Output Token Probabilities (Confidence)}
\label{sec:methods-output}
The simplest predictor uses the probability the model assigns to an answer in the
judging pass. It requires neither hidden states nor labeled training examples,
and we refer to it as \emph{confidence}. For program correctness, we use the
probability of \texttt{Yes}, $p_{\mathcal{M}}(\texttt{Yes} \mid x, c)$,
following 
Kadavath et
al.~\cite{Kadavath2022LanguageM}. For verdict correctness, we use the
probability of the emitted verdict, $p_{\mathcal{M}}(v(x, c) \mid x, c)$.
A model that is uncertain of its verdict should be
wrong more often~\cite{Hendrycks2017Baseline}. Both come from a softmax over the full vocabulary, whose
temperature is the only free parameter (Section~\ref{sec:setup-training}).

\subsection{Consistency (EigenScore)}
\label{sec:methods-consistency}
Consistency based methods do not judge a single candidate but asks how consistent
the model is across several generations for the same problem: a model that knows
how to solve a problem should produce similar solutions, while diverging
samples indicate uncertainty. 
We use EigenScore~\cite{Chen24Eigenscore},
which measures consistency in the hidden states. For a
problem with $m$ sampled candidates, it takes the hidden state of the last
layer at the last token of each candidate's generation and computes the
log-determinant of their regularized covariance matrix, normalized by $m$: the
more the states spread out, the higher this {\it Eigenscore}. We use the negated
EigenScore as an indicatore that 
instances from the candidate set solve the problem. 
EigenScore is an unsupervised method, requiring no correctness labels. Unlike all other methods, it scores a problem rather than a candidate and does not rank the candidates of one problem. We compare it with the other methods at problem level (Section~\ref{sec:rq-problem}). 
Its cost is $m$ generations per problem, which come for free wherever multiple candidates are sampled anyway.

\subsection{Internal Features}
\label{sec:methods-internal} 
The third family reads correctness from the hidden states. We extract the
hidden state $h_\ell(x, c) \in \mathbb{R}^d$ after a selected layer $\ell$ at the
last token of the judging prompt, the position from which the verdict is
generated. $h_\ell(x, c)$ serves as input for two probing methods:

\paragraph{Correctness probe}
We use linear probes~\cite{Alain2016Probes}, i.e. logistic regressors
trained on $h_\ell$ with correctness labels~\cite{OrgadTGRSKB25,
bui2025openia, Vu2026Autoprobe}: $y^{\mathrm{prog}}$ for program correctness and $y^{\mathrm{verd}}$ for verdict correctness. A probe's predicted probability of the positive class serves as the score. 
Since the classifier is linear, it captures only linearly accessible information~\cite{park2024linear}, like the other hidden states methods.

\paragraph{Linear Artificial Tomography (LAT)}
LAT~\cite{ribeiro2026repe, Zou2023RepE} identifies a correctness direction:
For each training problem with both a passing and a failing candidate, it computes the difference
of their hidden states and applies principal component analysis (PCA) to these difference vectors. 
The first principal component defines the direction, whose sign is chosen so that passing candidates tend to have a higher projection than failing ones. 
A candidate's score is its projection onto this direction.
In the original method, hidden states are extracted using a dedicated stimulus prompt that presents the problem and candidate~\cite{ribeiro2026repe}. In a preliminary comparison, applying the same procedure to the judging pass did not improve performance, so we use the original prompt for fairness. LAT was designed to predict program correctness, and we evaluate it for this target only.

\subsection{Internal Causal Features}
\label{sec:methods-causal}
These methods follow the tradition of causal interpretability, which identifies a
causal representation by intervening on it~\cite{Geiger2021CausalAbstractions}. 
Here, these approaches try to localize specific features within hidden states that causally determine a model's verdict rather than merely correlating with correctness. If a model judges code reliably when prompted, it presumably does so through a systematic internal
computation. 
We ask whether this computation can be localized 
and whether it helps predicting correctness on new inputs.

\begin{figure}[t]
\centering
\includegraphics{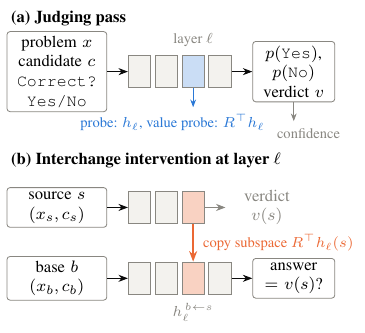}
\caption{(a) The judging pass: the model reads the problem, the candidate and
the question and answers \texttt{Yes} or \texttt{No} at the last prompt token.
The confidence baseline reads answer probabilities, the probe the hidden state
$h_\ell$ of one layer, the value probe its projection onto the learned
subspace. 
(b) An interchange intervention: the subspace coordinates of a
source example's hidden state replace those of a base example, and the base's forward pass continues. 
DAS learns a subspace so that the base then answers like the source.
Counterfactual simulation scores a candidate by making it the source for $B = 16$ fixed bases from the training set. A source state that firmly encodes its verdict should enforce it on the bases, one on which the model is unsure should not. The more consistently the bases adopt the candidate's verdict, the more likely it is correct. 
}
\label{fig:intervention}
\end{figure}

\paragraph{Localizing the verdict}
We assume that the model computes an internal assessment of whether a given
candidate is correct, a variable $V \in \{\texttt{Yes}, \texttt{No}\}$ that
determines its verdict and is encoded in the hidden state at the last prompt
token. To localize it, we use Distributed Alignment Search (DAS)~\cite{GeigerWPIG24DAS}, which can be understood by analogy to debugging: 
Run a program on two inputs, pause both at a breakpoint, copy a variable's value from the second run into the first, and continue: if the first run now produces the expected output of the second, the variable {\it controls} the output. 
The run that is modified is called the \emph{base} $b = (x_b, c_b)$, the run whose value is copied the \emph{source} $s = (x_s, c_s)$. 
Note that both are usually sampled from different problems (Figure~\ref{fig:intervention}). 

In an LLM there are no named variables whose values could be switched as in the debugging analogy, and a concept such as ``this program is correct'' is typically spread across many dimensions of $h_\ell$. 
DAS therefore \emph{learns} where to look: a $k$-dimensional subspace, given by $R \in \mathbb{R}^{d \times k}$ with orthonormal columns. 
The intervention replaces the  component of the base state's representation within this subspace with the corresponding component from the source
and leaves the rest of the base's state untouched:
\begin{equation}
  h_\ell^{\,b\leftarrow s} \;=\; h_\ell(b) + R R^{\top}\bigl(h_\ell(s) - h_\ell(b)\bigr).
\end{equation}
The forward pass of $b$ then continues through all later layers from
$h_\ell^{\,b\leftarrow s}$ instead of $h_\ell(b)$. Such a replacement is
called an interchange intervention~\cite{Geiger2021CausalAbstractions}.

\paragraph{Training}
DAS is trained only on candidates the model judges correctly ($y^{\mathrm{verd}}{=}1$). 
On these, the verdict coincides with program correctness $y^{\mathrm{prog}}$: \texttt{Yes} for the passing
candidates, \texttt{No} for the failing ones. 
A training pair combines two such candidates as base and source, from the same or different problems and with the same or different verdicts. 
With the LLM frozen, $R$ is trained so that the intervened base outputs the source's verdict: Where the two verdicts differ, the intervention must flip the base's answer; where they agree, it
must leave it unchanged. The share of held-out interventions that produce the
intended verdict is the interchange intervention accuracy (IIA)~\cite{GeigerWPIG24DAS}. 
It shows whether the subspace controls the verdict, not how well the verdict predicts correctness. We learn one subspace per layer and use it for both targets.

\paragraph{Counterfactual Simulation}
Counterfactual simulation~\cite{Huang2025Internal} asks whether a (source) candidate's
state in the subspace is strong enough to impose its assessment on other
(base) examples. The intuition here is that a source candidate whose state clearly encodes a verdict -- \texttt{Yes} or \texttt{No} --
should do so, while one on which the model is unsure should not. 
We fix a set $\mathcal{B}$ of $B {=} 16$ correctly judged training candidates as bases. 
The candidate to be scored is the source $s$. 
Let $p_{b \leftarrow s}(t)$ be the probability of answer $t$ in the intervened
forward pass of base $b$. Averaged over the bases, we calculate two scores:
\begin{align}
\mathrm{cs}_{\mathrm{prog}}(s) &= \frac{1}{B}\sum_{b \in \mathcal{B}}
p_{b \leftarrow s}(\texttt{Yes}), \\
\mathrm{cs}_{\mathrm{verd}}(s) &= \frac{1}{B}\sum_{b \in \mathcal{B}}
p_{b \leftarrow s}\bigl(v(s)\bigr).
\end{align}
The first measures how strongly the candidate's state pushes the bases' verdicts toward
\texttt{Yes} and predicts program correctness. 
The second measures how consistently the bases adopt the candidate's own verdict and predicts verdict
correctness. Note that counterfactual simulation requires $B$ additional forward passes per candidate.

\paragraph{Value Probe}
The value probe~\cite{Huang2025Internal} utilizes the subspace $R$ for probing instead of
intervening on it, so it needs no forward pass beyond the judging pass. Like the correctness probe (Section \ref{sec:methods-internal}), it uses logistic regression, but on
the $k$-dimensional projection $R^{\top} h_\ell(x, c)$ of the candidate's hidden state onto subspace $R$. 
This probe is trained differently for the two targets:
For the verdict target, it learns which verdict a subspace state encodes. It is trained on the same candidates as DAS, those the model judges correctly,
to predict the model's verdict, which on these candidates equals $y^{\mathrm{prog}}$: \texttt{Yes} for passing and \texttt{No} for failing programs. Following Huang et al., the
score is the probe's confidence in its own prediction, $|p(\texttt{Yes}) - \tfrac12|$. As in counterfactual simulation, a state that clearly reads as \texttt{Yes} or \texttt{No} indicates certainty and a probably correct verdict.
For the program target, we train the probe on all training candidates to predict $y^{\mathrm{prog}}$, and its probability of \texttt{Yes} is the score. 

\section{Study Design}
We evaluate the methods of Section~\ref{sec:methods} on both targets of
Section~\ref{sec:problem}, verdict and code correctness, in distribution and under benchmark and language
shift. RQ1--RQ3 judge single candidates, RQ4 whole problems.

\subsection{Models and Data}
\label{sec:setup-data}

\paragraph{Models}
We study two open-weight models, \texttt{Qwen3-Coder-30B-A3B-Instruct} and
OpenAI \texttt{gpt-oss-20b}. Each model generates candidate programs and judges its own candidates, so generation, verdicts and hidden states come from the same model. 
Qwen3-Coder answers the judging prompt directly. 
Since gpt-oss is a reasoning model, we prefill its final channel so that it also answers directly. 
Preliminary experiments showed that reasoning would cost about 1{,}100 generated tokens per candidate (median at medium effort) instead of a single forward pass, and its varying
length would shift the position at which the methods read the hidden state.
Judging without reasoning costs gpt-oss some accuracy (Section~\ref{sec:threats}).

\paragraph{Benchmarks}
We construct our datasets from LiveCodeBench~\cite{Jain2025Livecodebench} (\texttt{release\_v6}) and BigCodeBench~\cite{Zhuo2025BCB}, both written in Python. 
To evaluate  language shift we use 454 LiveCodeBench problems released between October 2024 and May 2025. We use the original Python versions, together with corresponding C++ versions from the LiveCodeBench-CPP dataset (\texttt{v6\_2408\_2505})\footnote{\url{https://huggingface.co/datasets/nvidia/LiveCodeBench-CPP}} and Java versions sourced from the original AtCoder and LeetCode platforms. We refer to these datasets as LCB-Python, LCB-C++, and LCB-Java, respectively. LCB-Python is the subset of our LCB data on these problems.

\paragraph{Candidates and labels}
For each problem, each model samples $n = 10$ candidates at temperature $t=1$. We discard generations from which no code candidate can be extracted, candidates whose judging prompts exceed 2048 tokens, and duplicate generations.
EigenScore compares $m$ non-empty candidates per problem: $m = 5$ on LCB and
the language sets, where some generations are empty, and all ten on BCB,
where almost none are. Problems with fewer non-empty candidates are left out
of the problem-level comparison for every method (93--100\% of the problems
are kept for Qwen3-Coder, 76--94\% for gpt-oss).

We obtain program-correctness labels by executing each candidate against the respective benchmark's test suite. To obtain verdict-correctness labels, we prompt the generating model to judge each candidate and compare its verdict with the execution-based correctness label
(Section~\ref{sec:problem}).

\paragraph{Splits}
We split each benchmark by problem into training, validation and test sets (60/20/20), so all candidates of a problem fall into the same split. 
For the language sets, with fewer than 90 test problems each, we evaluate on validation and test together. 
We repeat every experiment on three partitions of the
problems with disjoint test sets and report the mean over the partitions.

Table~\ref{tab:datasets} summarizes the evaluation sets of the first
partition and reports the fraction of candidates that pass the tests, the
fraction accepted by the model, and the balanced accuracy of the model's
verdict against program correctness. The models' balanced accuracy ranges from
57.2\% to 74.4\%, so the models misjudge many of their own candidates, and
knowing which verdicts to trust matters.

\begin{table}[t]
\centering
\caption{Evaluation sets of the first partition. \emph{Problems}: number of distinct problems; \emph{Cand.}: number
of scored candidates; \emph{Pass}: percentage of candidates that pass the tests; \emph{Yes}: percentage of candidates accepted by the model;
\emph{Judge acc.}: balanced accuracy of the model's verdict against program correctness (mean of the acceptance rate on passing and the rejection rate on failing candidates). }
\label{tab:datasets}
\begin{tabular}{l rrrrr}
\toprule
Benchmark & Problems & Cand. & Pass & Yes & Judge acc. \\
\midrule
\multicolumn{6}{l}{\emph{Qwen3-Coder-30B}} \\
LiveCodeBench & 208 & 1646 & 55.3 & 58.3 & 74.4 \\
BigCodeBench  & 456 & 3240 & 40.2 & 71.7 & 58.3 \\
LCB-Python    & 178 & 1423 & 41.7 & 44.1 & 73.8 \\
LCB-C++       & 175 & 1448 & 24.0 & 32.5 & 63.8 \\
LCB-Java      & 179 & 1486 & 37.5 & 31.2 & 64.4 \\
\midrule
\multicolumn{6}{l}{\emph{gpt-oss-20b}} \\
LiveCodeBench & 189 & 1588 & 80.4 & 44.6 & 62.8 \\
BigCodeBench  & 456 & 4533 & 45.8 & 42.6 & 61.7 \\
LCB-Python    & 150 & 1140 & 76.8 & 37.6 & 62.2 \\
LCB-C++       & 147 & 1116 & 68.4 & 35.1 & 57.2 \\
LCB-Java      & 141 & 1069 & 59.8 & 38.6 & 65.4 \\
\bottomrule
\end{tabular}
\end{table}

\subsection{Training and Selection}
\label{sec:setup-training}
\paragraph{Fitting}
The probe is a logistic regression on the hidden state of one layer, trained
on 512 candidates per class. LAT uses the stimulus prompt of Ribeiro et
al.~\cite{ribeiro2026repe} and is fitted on one pair of a passing and a
failing candidate per training problem that has both. DAS learns a
$k {=} 6$-dimensional subspace per layer, a size chosen in a preliminary sweep
and in line with evidence that correctness is encoded in few
dimensions~\cite{cho2026confidencemanifold}. It is trained on pairs of 1024
correctly judged training candidates, in at most one pass over 20,480 pairs
with early stopping on the training IIA. The value probe is a logistic
regression on the projections of the same training candidates onto the subspace.
Counterfactual simulation uses a fixed set of $B {=} 16$ correctly judged
training candidates as bases. The generation probes of RQ3 apply the probe to
the hidden states of the model's own generation, replayed token by token, at
the last token of the code or of the prompt; they are trained on the same
candidates and program labels as the probe. Detailed settings can be found in the replication package\footnote{\url{https://anonymous.4open.science/r/FORGE-PCCuDS}}.

\paragraph{Selection}
Each method comes with one free parameter: the layer, or --
for confidence -- the temperature. The probe and LAT are fitted on every layer, while DAS
is trained at a subset (every second layer of L33--L43 for Qwen3-Coder and of L11--L23 for gpt-oss), since each subspace takes several GPU hours.
Layers are chosen in distribution, via cross-fitting over problems.
The selection is made on four fifths of the evaluation problems and read on the
remaining fifth, over 20 random splits. Under shift, the choice made on all
in-distribution rows is read on the shifted set, so no shifted row enters
selection. 

\subsection{Evaluation Protocol}
\label{sec:setup-protocol}
\paragraph{Metrics}

At candidate level (RQ1--RQ3) we report the area under the receiver operating characteristic curve (AUROC)
over all candidates in an evaluation set, as is common for correctness prediction~\cite{he2026usingsemanticdistanceestimate,he2026codecircuit,bouchard2026functional,shi2026codemorethantext}, using either verdict correctness or program correctness as the positive label. AUROC measures how well a score ranks positive examples above negative ones, with 0.5 corresponding to chance and 1.0 to perfect ranking.
Because candidates from different problems are pooled, this metric reflects both the ability to distinguish candidates within a problem and the ability to rank problems by how often the model produces positive outcomes. Where this distinction matters, we additionally report within-problem AUROC, computed over pairs of candidates from the same problem with different labels. It equals the pairwise accuracy of~\cite{ribeiro2026robustness} and is on the same scale as the pooled AUROC.
The gap between the two is the part of the pooled AUROC that comes from ranking problems.
At problem level (RQ4), a problem is
considered as solved if at least half of its candidates pass, and the
single-pass methods are averaged over the candidates that EigenScore uses.

\paragraph{Significance}

We compare methods using a paired, problem-clustered bootstrap with 2,000 resamples. Each resample draws whole problems and evaluates all methods on the same examples, accounting for within-problem dependence while preserving paired comparisons. For language-shift results, we average differences over the six language pairs and three partitions, using shared resamples across language pairs because they share problems. We combine the bootstrap standard error (SE) with between-partition variability to account for variation due to training and layer selection. We consider a difference significant if $|\Delta| > 1.96\,\mathrm{SE}$ and report $\Delta \pm 1.96\,\mathrm{SE}$.


\section{Experiments}

\subsection{RQ1: How well can execution-free methods tell whether the model's own verdict can be trusted? }
\label{sec:rq-verdict}

\begin{table}[t]
\centering\footnotesize
\setlength{\tabcolsep}{3.2pt}
\caption{AUROC for verdict correctness (mean over 3 partitions). \emph{LCB}, \emph{BCB}: trained and evaluated on that benchmark; $\to$: evaluated on the other one. Language shift: \emph{ID} is the mean over the three training languages, \emph{Shift} the mean over the six OOD train$\to$evaluation pairs (Table~\ref{tab:language-detail-verdict}), $\Delta$ their difference. 
Best per column and LLM in bold.}
\label{tab:verdict-summary}
\begin{tabular}{l rrrr rrr}
\toprule
 & \multicolumn{4}{c}{Benchmark shift} & \multicolumn{3}{c}{Language shift} \\
\cmidrule(lr){2-5}\cmidrule(lr){6-8}
Method & LCB & $\to$BCB & BCB & $\to$LCB & ID & Shift & $\Delta$ \\
\midrule
\multicolumn{8}{l}{\emph{Qwen3-Coder-30B}} \\
Confidence & .672 & \textbf{.552} & .555 & .669 & .644 & .643 & \textbf{$-$.001} \\
Probe & .699 & .480 & \textbf{.639} & .516 & \textbf{.761} & .668 & $-$.093 \\
\addlinespace[2pt]
Counterf.\ sim. & .692 & .453 & .560 & .680 & .759 & \textbf{.728} & $-$.031 \\
Value probe & \textbf{.740} & .485 & .579 & \textbf{.695} & .758 & .728 & $-$.031 \\
\midrule
\multicolumn{8}{l}{\emph{gpt-oss-20b}} \\
Confidence & .569 & .598 & .597 & .544 & .542 & .536 & \textbf{$-$.005} \\
Probe & \textbf{.829} & .478 & .507 & .476 & .774 & .668 & $-$.106 \\
\addlinespace[2pt]
Counterf.\ sim. & .779 & .572 & \textbf{.601} & .556 & \textbf{.808} & \textbf{.691} & $-$.116 \\
Value probe & .695 & \textbf{.599} & .600 & \textbf{.605} & .713 & .646 & $-$.067 \\
\bottomrule
\end{tabular}
\end{table}

\begin{table*}[t]
\centering\footnotesize
\setlength{\tabcolsep}{1.5pt}
\caption{AUROC for verdict correctness, in distribution (ID) and under language shift per training language (LCB problems; ID cross-fitted, shifted columns at the layer chosen in distribution). Mean $\pm$ SD over 3 partitions of the problems. Best per column in bold; underlined: within one SD of the best. Non-causal methods above the rule, causal below.}
\label{tab:language-detail-verdict}
\begin{tabular*}{\textwidth}{@{\extracolsep{\fill}}ll rrr rrr rrr@{}}
\toprule
 & & \multicolumn{3}{c}{Trained on Py} & \multicolumn{3}{c}{Trained on C++} & \multicolumn{3}{c}{Trained on Java} \\
\cmidrule(lr){3-5}\cmidrule(lr){6-8}\cmidrule(lr){9-11}
Model & Method & ID & $\to$C++ & $\to$Java & ID & $\to$Py & $\to$Java & ID & $\to$Py & $\to$C++ \\
\midrule
\multirow{4}{*}{Qwen3-Coder-30B} & Confidence & .608\,{\scriptsize$\pm$.028} & .637\,{\scriptsize$\pm$.052} & .687\,{\scriptsize$\pm$.011} & .637\,{\scriptsize$\pm$.054} & .611\,{\scriptsize$\pm$.026} & .688\,{\scriptsize$\pm$.013} & .689\,{\scriptsize$\pm$.010} & .605\,{\scriptsize$\pm$.027} & .630\,{\scriptsize$\pm$.053} \\
 & Probe & .670\,{\scriptsize$\pm$.011} & \underline{.628}\,{\scriptsize$\pm$.103} & .636\,{\scriptsize$\pm$.059} & \textbf{.874}\,{\scriptsize$\pm$.010} & .636\,{\scriptsize$\pm$.062} & .701\,{\scriptsize$\pm$.009} & \underline{.740}\,{\scriptsize$\pm$.044} & .670\,{\scriptsize$\pm$.026} & .739\,{\scriptsize$\pm$.020} \\
\cmidrule(lr){3-5}\cmidrule(lr){6-8}\cmidrule(lr){9-11}
 & Counterf.\ sim. & \underline{.671}\,{\scriptsize$\pm$.071} & \underline{.679}\,{\scriptsize$\pm$.129} & \underline{.721}\,{\scriptsize$\pm$.041} & \underline{.854}\,{\scriptsize$\pm$.021} & \underline{.714}\,{\scriptsize$\pm$.024} & .746\,{\scriptsize$\pm$.013} & \textbf{.751}\,{\scriptsize$\pm$.019} & \underline{.729}\,{\scriptsize$\pm$.017} & \textbf{.778}\,{\scriptsize$\pm$.013} \\
 & Value probe & \textbf{.717}\,{\scriptsize$\pm$.031} & \textbf{.692}\,{\scriptsize$\pm$.030} & \textbf{.737}\,{\scriptsize$\pm$.004} & .814\,{\scriptsize$\pm$.010} & \textbf{.731}\,{\scriptsize$\pm$.003} & \textbf{.761}\,{\scriptsize$\pm$.010} & \underline{.744}\,{\scriptsize$\pm$.011} & \textbf{.733}\,{\scriptsize$\pm$.019} & .712\,{\scriptsize$\pm$.026} \\
\midrule
\multirow{4}{*}{gpt-oss-20b} & Confidence & .542\,{\scriptsize$\pm$.023} & .529\,{\scriptsize$\pm$.016} & .549\,{\scriptsize$\pm$.050} & .528\,{\scriptsize$\pm$.016} & .541\,{\scriptsize$\pm$.023} & .549\,{\scriptsize$\pm$.050} & .555\,{\scriptsize$\pm$.045} & .525\,{\scriptsize$\pm$.009} & .525\,{\scriptsize$\pm$.016} \\
 & Probe & \underline{.812}\,{\scriptsize$\pm$.021} & \textbf{.692}\,{\scriptsize$\pm$.016} & .677\,{\scriptsize$\pm$.018} & \underline{.797}\,{\scriptsize$\pm$.011} & \underline{.651}\,{\scriptsize$\pm$.045} & .596\,{\scriptsize$\pm$.021} & .713\,{\scriptsize$\pm$.033} & \underline{.739}\,{\scriptsize$\pm$.055} & \underline{.651}\,{\scriptsize$\pm$.018} \\
\cmidrule(lr){3-5}\cmidrule(lr){6-8}\cmidrule(lr){9-11}
 & Counterf.\ sim. & \textbf{.828}\,{\scriptsize$\pm$.099} & \underline{.652}\,{\scriptsize$\pm$.144} & \textbf{.759}\,{\scriptsize$\pm$.034} & \textbf{.799}\,{\scriptsize$\pm$.088} & \textbf{.670}\,{\scriptsize$\pm$.168} & \textbf{.624}\,{\scriptsize$\pm$.057} & \textbf{.796}\,{\scriptsize$\pm$.027} & \textbf{.784}\,{\scriptsize$\pm$.092} & \textbf{.660}\,{\scriptsize$\pm$.091} \\
 & Value probe & .743\,{\scriptsize$\pm$.010} & .636\,{\scriptsize$\pm$.028} & .645\,{\scriptsize$\pm$.028} & .700\,{\scriptsize$\pm$.007} & .633\,{\scriptsize$\pm$.023} & .591\,{\scriptsize$\pm$.007} & .697\,{\scriptsize$\pm$.005} & .721\,{\scriptsize$\pm$.018} & \underline{.653}\,{\scriptsize$\pm$.015} \\
\bottomrule
\end{tabular*}
\end{table*}


Both LLMs are mediocre judges, with balanced accuracies of their verdicts of 57--74\% (Table~\ref{tab:datasets}) against the tests. RQ1 asks how well our methods identify the candidates on which the model's verdict is correct.
Table~\ref{tab:verdict-summary} summarizes the results, showing the mean per method over partitions.
Table~\ref{tab:language-detail-verdict} breaks down the language shift by training and evaluation language.

\paragraph{In distribution, hidden states outperform confidence}
On LCB and the language sets, the methods that read hidden states reach 0.69--0.83, while confidence is weak for
Qwen3-Coder (0.64--0.67) and close to chance for gpt-oss (0.54--0.57). 
On the language sets, the probe and the two causal methods are
level for Qwen3-Coder (0.758--0.761); for gpt-oss, counterfactual simulation is best (0.808). BCB is harder, no trained method exceeds 0.639 for Qwen3-Coder or 0.601 for gpt-oss.

\begin{figure}[t]
\centering
\includegraphics[width=\columnwidth]{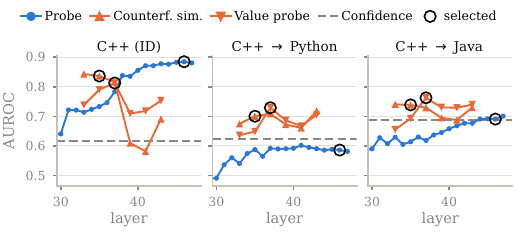}
\caption{AUROC per layer on the verdict target, Qwen3-Coder trained on C++
(fold 0), in distribution (left) and under language shift. Rings mark the
layer chosen in distribution; the dashed line is confidence. Under shift to
Python, every causal checkpoint lies above the probe's best layer; on Java,
most do.}
\label{fig:verdict-profile}
\end{figure}

\paragraph{Under language shift, the causal subspace is the most robust predictor for Qwen3-Coder}
For Qwen3-Coder, the probe drops from 0.76 in distribution to 0.67 when the language changes
, while both causal methods fall only from 0.76 to 0.73, against 0.64 for confidence. Averaged over the six language pairs, the value probe is ahead of
the best non-causal method of each pair by $0.042 \pm 0.021$ (95\% interval of the paired test, Section~\ref{sec:setup-protocol}) and ahead in each of the three partitions ($+0.035$, $+0.053$, $+0.038$). Counterfactual simulation
reaches the same mean but varies more between partitions ($\pm 0.043$). 
The lead does not depend on the probe's layer choice: in 13 of the 18 pair--partition combinations, most value-probe checkpoints lie above the probe at its best layer on the shifted rows, as illustrated for Qwen3-Coder, trained on C++ in Figure~\ref{fig:verdict-profile}. 
For gpt-oss, all trained methods lose 0.07--0.12 AUROC under language shift, and neither causal method is significantly ahead of the best non-causal one. 
The robustness reported for natural language tasks~\cite{Huang2025Internal} thus carries over to code for one of the two models; Section~\ref{sec:rq-subspace} examines what carries the gain.

\begin{figure}[t]
\centering
\includegraphics[width=\columnwidth]{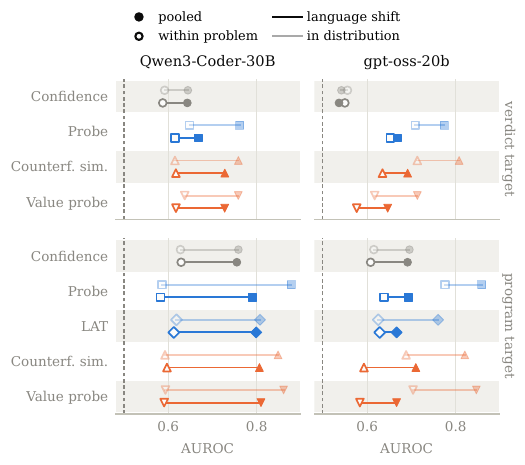}
\caption{Within-problem (hollow) against pooled AUROC (filled), in distribution
(faint) and under language shift: means over training languages and partitions.
Within-problem AUROC uses the problems with mixed outcomes only
(Section~\ref{sec:setup-protocol}); the length of a line is how much of the
pooled AUROC comes from ranking problems.}
\label{fig:within}
\end{figure}

\paragraph{The advantage lies in identifying unreliable problems, not candidates}
Pooled AUROC combines the ability to identify problems on which the judge is unreliable with the ability to distinguish correct from incorrect verdicts within a problem (Section~\ref{sec:setup-protocol}). The second is limited. Under language shift, within-problem AUROC is 0.616--0.618 for all trained methods of Qwen3-Coder. 
For gpt-oss the probe reaches 0.652, counterfactual simulation 0.635 and the value probe 0.576 (Figure~\ref{fig:within}).
The causal methods' pooled advantage thus comes mainly from identifying problems on which the judge is unreliable, rather than individual candidates it misjudges. 

\paragraph{A change of benchmark affects every method more than a change of
language}
Transfer between LCB 
and BCB 
is asymmetric (Table~\ref{tab:verdict-summary}). From LCB to BCB, all trained methods of Qwen3-Coder fall
to 0.453--0.485, below confidence (0.552). In the reverse direction, the probe stays near chance (0.516), while the causal methods reach 0.680 and 0.695, close to confidence (0.669).
For gpt-oss, every method stays at 0.48--0.61 in both directions. 

\begin{rqanswer}{RQ1}
In distribution, hidden states predict whether the model's verdict is right far
better than its confidence. Under language shift, a probe on the causally
localized subspace is the most robust predictor for Qwen3-Coder (+0.04 AUROC over
the best non-causal method, in all three partitions), mainly by identifying problems
on which the judge is unreliable; for gpt-oss, no method is significantly ahead. 
Verdict prediction does not transfer reliably between benchmarks.
\end{rqanswer}

\subsection{RQ2: Which part of the causal methods carries their signal?}
\label{sec:rq-subspace}

On Qwen3-Coder's verdict target, the causal methods are the most robust predictors under language shift (RQ1). 
We now investigate which aspects of the method contribute to this performance: whether prediction requires the intervention itself, whether the learned directions matter beyond their dimensionality, and whether interchange intervention accuracy (see Section~\ref{sec:methods-causal}) indicates when the subspace is useful.

\begin{figure}[t]
\centering
\includegraphics[width=\columnwidth]{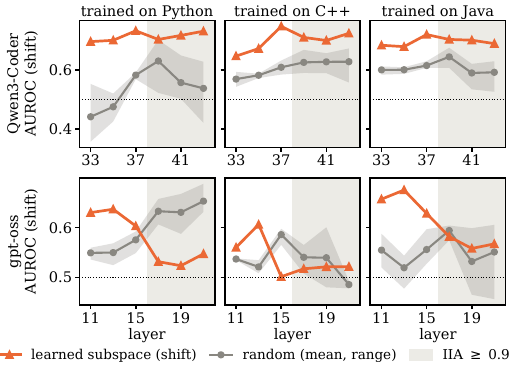}
\caption{Value probe on the learned DAS subspace against value probes on
three random 6-dimensional subspaces at the same layer, verdict target,
mean over shifted languages (the band is the range over
the random subspaces). Shaded: layers where DAS has
converged (IIA $\geq$ 0.9). On Qwen3-Coder the learned subspace is above
random at every layer; on gpt-oss it is above random before convergence and
near chance once DAS has converged.}
\label{fig:subspace-layers}
\end{figure}

\paragraph{Reading the subspace is sufficient}
Counterfactual simulation intervenes through the subspace and reads the answer
token, at the cost of $B$ additional forward passes per candidate; the value
probe reads the subspace directly in the judging pass. On Qwen3-Coder the two are equally robust under language shift (mean difference over the six pairs $-0.000 \pm 0.033$ on the verdict target, $+0.004 \pm 0.007$ on the program target), and both are ahead of confidence, which reads the same
answer token without the subspace ($+0.085 \pm 0.033$ and $+0.05 \pm 0.03$).
The intervention thus brings no significant improvement over reading the
subspace, and the value probe costs no more at prediction time than a probe.

\paragraph{The learned direction matters, not the low dimension}
A six-dimensional readout might transfer better simply because it is small. We
therefore train the value probe on three random subspaces of the same
dimension at the same layer. At the layer chosen in distribution, the
learned subspace reaches 0.70--0.75 under shift for Qwen3-Coder, random
subspaces 0.48--0.61, no better than the probe on the full hidden state (0.668,
Table~\ref{tab:verdict-summary}). The same holds at every trained layer (Figure~\ref{fig:subspace-layers}). 


\paragraph{Interchange accuracy does not guarantee predictive utility}
DAS learns the subspace so that swapping it between two inputs swaps the
model's verdict. IIA measures how often this succeeds. A high IIA thus shows
that the subspace controls the decision, not that it tells whether the decision
is right. On gpt-oss the two diverge (Figure~\ref{fig:subspace-layers}): at L11
and L13, where DAS succeeds only partly (IIA 0.54--0.64), the learned subspace
transfers better than random ones (0.56--0.68 against 0.52--0.55 under shift),
and L13 is the layer chosen in distribution. However, at L17--L21, where DAS almost
always succeeds (IIA 0.91--0.95), it falls to 0.52--0.58, no better than
random. 
On Qwen3-Coder the converged layers (L39--L43, IIA 0.95) remain informative. IIA thus cannot tell the two cases apart, and layers are chosen by held-out prediction instead (Section~\ref{sec:setup-training}).

\begin{rqanswer}{RQ2}
Where the causal methods help, the learned subspace carries the gain: the value
probe, which reads it in the judging pass, is as robust as counterfactual
simulation, and random subspaces of the same size transfer only as well as the
full hidden state. 
IIA does not indicate when the subspace is useful: on gpt-oss, the
converged subspaces are no better than random ones.
\end{rqanswer}

\subsection{RQ3: How well do the methods predict whether the program is correct?}
\label{sec:rq-program}

The program target asks whether a candidate passes its tests, regardless of
the model's verdict. The two targets
agree only where the judge is right. 
We therefore ask whether the results on the verdict target carry over. Table~\ref{tab:program-summary} summarizes the results. In distribution, all methods predict
program correctness well above chance, AUROC ranges from 0.70--0.88 on LCB and the language
sets, with the hidden-state probe performing best on the language sets for both models. BCB is more challenging for every method.

\begin{table}[t]
\centering\footnotesize
\setlength{\tabcolsep}{3.2pt}
\caption{AUROC for program correctness (mean over 3 partitions). \emph{LCB}, \emph{BCB}: trained and evaluated on that benchmark; $\to$: evaluated on the other one. Language shift: \emph{ID} is the mean over the three training languages, \emph{Shift} the mean over the six train$\to$evaluation pairs, $\Delta$ their difference. 
Best per column and model in bold.}
\label{tab:program-summary}
\begin{tabular}{l rrrr rrr}
\toprule
 & \multicolumn{4}{c}{Benchmark shift} & \multicolumn{3}{c}{Language shift} \\
\cmidrule(lr){2-5}\cmidrule(lr){6-8}
Method & LCB & $\to$BCB & BCB & $\to$LCB & ID & Shift & $\Delta$ \\
\midrule
\multicolumn{8}{l}{\emph{Qwen3-Coder-30B}} \\
Confidence & .815 & .650 & .648 & .815 & .759 & .755 & \textbf{$-$.003} \\
Probe & \textbf{.844} & \textbf{.656} & \textbf{.688} & .697 & \textbf{.878} & .790 & $-$.088 \\
LAT & .844 & .580 & .590 & .731 & .807 & .799 & $-$.009 \\
\addlinespace[2pt]
Counterf.\ sim. & .839 & .644 & .635 & .813 & .848 & .806 & $-$.042 \\
Value probe & .838 & .642 & .664 & \textbf{.815} & .861 & \textbf{.810} & $-$.051 \\
\midrule
\multicolumn{8}{l}{\emph{gpt-oss-20b}} \\
Confidence & .723 & \textbf{.648} & .646 & .721 & .696 & .691 & \textbf{$-$.004} \\
Probe & .793 & .575 & .649 & .690 & \textbf{.859} & .693 & $-$.166 \\
LAT & .767 & .618 & .608 & .676 & .760 & .666 & $-$.094 \\
\addlinespace[2pt]
Counterf.\ sim. & .780 & .647 & .652 & \textbf{.724} & .820 & \textbf{.710} & $-$.111 \\
Value probe & \textbf{.803} & .638 & \textbf{.652} & .712 & .846 & .666 & $-$.180 \\
\bottomrule
\end{tabular}
\end{table}

\paragraph{Under language shift, the causal methods lose their advantage}
The full-state probe loses 0.088 AUROC for Qwen3-Coder and 0.166 for gpt-oss, while confidence changes little ($-0.003$ and $-0.004$). For Qwen3-Coder, LAT and both causal methods nevertheless remain significantly ahead of confidence by 0.044--0.054 AUROC on average; the probe's lead of 0.035 is no longer significant. For gpt-oss, no trained method significantly outperforms confidence. Importantly, neither causal method outperforms the best non-causal method: for Qwen3-Coder, the value probe and counterfactual simulation trail it by 0.010 and 0.014 AUROC, respectively, neither significantly. For gpt-oss, the value probe is significantly behind ($-0.068$), while counterfactual simulation is statistically level ($-0.025$). Thus, the robustness of causal localization observed for verdict correctness (RQ1) does not carry over to program correctness.



\paragraph{Program correctness transfers between benchmarks}
In contrast to verdict correctness, predictors for program correctness transfer between LCB and BCB in both directions. When trained on LCB and evaluated on BCB, methods reach 0.580--0.656 for Qwen3-Coder, and 0.575--0.648 for gpt-oss, comparable to
the respective in-distribution results on BCB (0.590--0.688 and 0.608--0.652).
 Transfer in the reverse direction is similarly strong, reaching 0.70--0.82 for Qwen3-Coder and 0.68--0.72 for gpt-oss.
\paragraph{Pooled AUROC overstates candidate-level discrimination}
As for verdict correctness, pooled AUROC reflects both the ability to identify problems the model tends to solve and the ability to distinguish candidates within a problem.
Choosing a candidate needs only the second, and here every method is weak (Figure~\ref{fig:within}, bottom): under language shift, within-problem AUROC lies at 0.58--0.64 for all methods and both models, against a pooled AUROC of 0.67--0.81. For Qwen3-Coder confidence performs best within problem (0.629). 
The trained methods reach higher within-problem AUROCs of 0.69--0.78 only for gpt-oss in distribution, but lose this advantage under shift. Consequently, the predictors are more useful for identifying problems that warrant verification than for choosing among candidates of the same problem.
Section~\ref{sec:rq-problem} predicts at problem level directly.

\paragraph{Reading the generation is as good as asking}
All methods so far read the model in a separate judging pass. The same information might already be present when the model generates the candidate. 
We therefore train the same probe on the hidden states of the model's own
generation (without the judgment prompt), replayed token by token, and compare it with the judging-pass probe on identical candidates, Table~\ref{tab:genprobe} reports the results. We compare different token positions here: at the last token of the candidate's code, it is
at least as good as the judging pass in distribution (0.895 against 0.886 for
Qwen3-Coder, 0.864 against 0.851 for gpt-oss); under language shift, it is
close for Qwen3-Coder (0.795 against 0.779) and behind for gpt-oss (0.673
against 0.704). A correctness score can thus be read while the code is
written. Read at the last prompt token, before any code exists, the probe
has seen only the problem, yet for Qwen3-Coder it matches the judging pass in
distribution (0.891). Much of what the methods predict is thus the problem,
not the candidate.

\begin{table}[t]
\centering\footnotesize
\setlength{\tabcolsep}{3.2pt}
\caption{The probe on the judging pass against the same probe on the states of the model's own generation, program target on the language sets. \emph{ID}: mean AUROC over the three languages, \emph{Shift}: mean over the six language pairs, \emph{Within}: AUROC over pairs of candidates of the same problem. The last prompt token has seen only the problem.}
\label{tab:genprobe}
\begin{tabular}{l rr rr}
\toprule
 & \multicolumn{2}{c}{ID} & \multicolumn{2}{c}{Lang. Shift} \\
\cmidrule(lr){2-3}\cmidrule(lr){4-5}
Readout & Pooled & Within & Pooled & Within \\
\midrule
\multicolumn{5}{l}{\emph{Qwen3-Coder-30B}} \\
Judging pass & .886 & .553 & .779 & .540 \\
Generation, last code token & .895 & .620 & .795 & .627 \\
Generation, last prompt token & .891 & .514 & .755 & .531 \\
\midrule
\multicolumn{5}{l}{\emph{gpt-oss-20b}} \\
Judging pass & .851 & .772 & .704 & .609 \\
Generation, last code token & .864 & .781 & .673 & .623 \\
Generation, last prompt token & .720 & .524 & .669 & .478 \\
\bottomrule
\end{tabular}
\end{table}

\begin{rqanswer}{RQ3}
All methods predict program correctness well above chance, largely by recognizing the problems the model fails on, 
and this transfers better between benchmarks than the verdict. Under
language shift, the trained methods lose most of their lead over confidence:
about 0.05 remains for Qwen3-Coder, nothing for gpt-oss, and the causal
methods' robustness on the verdict target does not carry over. A probe on the model's own generation is as good as one on a separate judging pass in distribution.
\end{rqanswer}

\subsection{RQ4: Does agreement across samples improve problem-level prediction?} 
\label{sec:rq-problem}

This question moves from the candidate to the problem level. Much of the
candidate-level signal reflects which problems the model solves
(Section~\ref{sec:rq-program}). When several candidates are sampled for a
problem, their agreement provides a further signal.
EigenScore~\cite{Chen24Eigenscore} reads this agreement from the hidden states
of the generations (Section~\ref{sec:methods-consistency}). Like confidence, it
needs no training, so it cannot lose under shift. We compare at equal sampling
cost: each single-pass method is averaged over the candidates EigenScore uses,
and every method predicts whether at least half of a problem's candidates pass
(Table~\ref{tab:problem-level}).



\begin{table}[t]
\centering\footnotesize
\setlength{\tabcolsep}{3.2pt}
\caption{AUROC for whether a problem is solved (at least half of its candidates pass). Single-candidate methods are averaged over the candidates of a problem, at the layer of Table~\ref{tab:program-summary}; EigenScore is computed from the sampled candidates and needs no training. Columns as in Table~\ref{tab:program-summary}. $\uparrow$: significantly above EigenScore (paired problem bootstrap, 95\%; for ID and Shift on the mean over languages). All methods on the problems that have an EigenScore.$^{*}$}
\label{tab:problem-level}
\begin{tabular}{l rrrr rr}
\toprule
 & \multicolumn{4}{c}{Benchmark shift} & \multicolumn{2}{c}{Language shift} \\
\cmidrule(lr){2-5}\cmidrule(lr){6-7}
Method & LCB & $\to$BCB & BCB & $\to$LCB & ID & Shift \\
\midrule
\multicolumn{7}{l}{\emph{Qwen3-Coder-30B}} \\
EigenScore & .504 & .658 & .658 & .504 & .528 & .528 \\
Confidence & .877$^{\uparrow}$ & \textbf{.670} & .671 & .872$^{\uparrow}$ & .774$^{\uparrow}$ & .775$^{\uparrow}$ \\
Probe & .906$^{\uparrow}$ & .638 & .671 & .884$^{\uparrow}$ & \textbf{.911}$^{\uparrow}$ & .803$^{\uparrow}$ \\
LAT & .902$^{\uparrow}$ & .618 & \textbf{.682} & .871$^{\uparrow}$ & .844$^{\uparrow}$ & .821$^{\uparrow}$ \\
\addlinespace[2pt]
Counterf.\ sim. & \textbf{.912}$^{\uparrow}$ & .646 & .676 & .881$^{\uparrow}$ & .875$^{\uparrow}$ & .831$^{\uparrow}$ \\
Value probe & .904$^{\uparrow}$ & .652 & .680 & \textbf{.889}$^{\uparrow}$ & .890$^{\uparrow}$ & \textbf{.838}$^{\uparrow}$ \\
\midrule
\multicolumn{7}{l}{\emph{gpt-oss-20b}} \\
EigenScore & .660 & .518 & .518 & .660 & .586 & .586 \\
Confidence & .698 & \textbf{.677}$^{\uparrow}$ & .678$^{\uparrow}$ & .692 & .623 & .615 \\
Probe & .791 & .615 & \textbf{.712}$^{\uparrow}$ & \textbf{.759} & \textbf{.845}$^{\uparrow}$ & \textbf{.676}$^{\uparrow}$ \\
LAT & \textbf{.808}$^{\uparrow}$ & .662$^{\uparrow}$ & .622$^{\uparrow}$ & .688 & .772$^{\uparrow}$ & .645 \\
\addlinespace[2pt]
Counterf.\ sim. & .772 & .655$^{\uparrow}$ & .685$^{\uparrow}$ & .692 & .800$^{\uparrow}$ & .662 \\
Value probe & .790 & .663$^{\uparrow}$ & .692$^{\uparrow}$ & .676 & .834$^{\uparrow}$ & .630 \\
\bottomrule
\end{tabular}
\par\smallskip
\parbox{\linewidth}{\scriptsize $^{*}$EigenScore needs $m$ non-empty samples: five per problem on LCB and the language sets, ten on BCB. Problems with fewer non-empty generations are left out for every method, and the single-candidate methods are averaged over the candidates of the same samples. This keeps 93--100\% of the evaluation problems for Qwen3-Coder-30B and 76--94\% for gpt-oss-20b. 
}
\end{table}

The cheapest prediction needs no sample at all: for Qwen3-Coder, a probe that
has seen only the problem 
already ranks candidates as well as the judging pass
(Section~\ref{sec:rq-program}). Sampling several candidates pays off for
prediction only if their agreement adds to what a single pass provides.

\paragraph{Agreement provides no consistent advantage}
For Qwen3-Coder, EigenScore is at chance on LCB (0.504) and on the language
sets (0.528), where every single-candidate method, confidence included, is
significantly ahead (0.77--0.91). On BCB it is comparable to them (0.658 against
0.671--0.682). For gpt-oss the benchmarks swap roles: EigenScore is at chance
on BCB (0.518 against 0.622--0.712) and closer on LCB (0.660 against
0.698--0.808). Under shift, the single-pass methods lose ground. 
From LCB to BCB, Qwen3-Coder's trained methods fall slightly below EigenScore (0.618--0.652 against 0.658) while confidence stays above it (0.670); under language shift, the
gap remains large for Qwen3-Coder (0.775--0.838 against 0.528) and narrows for
gpt-oss (0.615--0.676 against 0.586).

\begin{rqanswer}{RQ4}
Agreement across samples gives no consistent advantage over a single pass.
EigenScore is level with the single-pass methods in some cells and near
chance in others, and confidence, which needs no labels either, is never
significantly below it. That EigenScore cannot lose under shift does not help where it starts near chance.
\end{rqanswer}


\section{Discussion}
\label{sec:discussion}

\paragraph{Verdict and program are different prediction targets}
Our results show that the usefulness of execution-free correctness prediction depends on both the prediction target and the type of distribution shift. In particular, predicting whether a model's verdict is correct is different from predicting whether the program itself is correct, even though both targets use signals from the same judging pass.
The two targets respond differently to shift. Causal localization helps only on the verdict, and between benchmarks only program correctness transfers. An execution-free correctness predictor for an LLM-based code-generation system should therefore be chosen and evaluated for the decision it supports. To decide which candidates to test in a new programming language, output confidence is a strong baseline: it needs no labels and loses almost nothing under shift,
while trained predictors keep only part of their lead. To decide whether to
trust the model's verdict, confidence is weak, and no verdict predictor we tested
carries over to a new benchmark. 


\paragraph{Triage rather than selection}
Much of what the methods predict is which problems
the model solves: every method is weak at ranking candidates within a problem, and a probe that has seen only the problem matches one that has seen the candidate (Table \ref{tab:genprobe}). Pooled metrics,  such as AUROC~\cite{bouchard2026functional, shi2026codemorethantext} or accuracy and F1~\cite{bui2025openia, Vu2026Autoprobe}, thus mix problem difficulty with the ability to tell a correct candidate from an incorrect one. 
For triage across problems this is useful, while for choosing among
the candidates of one problem it is not. Selection metrics~\cite{ribeiro2026repe} and pairwise
accuracy within a task~\cite{ribeiro2026robustness} measure the second ability
directly; we recommend reporting such a within-problem measure alongside the
pooled one. 

\paragraph{Causal localization for code}
Huang et al.~\cite{Huang2025Internal} report that causally localized
predictors are more robust under shift on natural-language tasks. For code,
this holds on one of our two models and for the target the method was
designed for, the model's own verdict; on the program correctness target it gives no
advantage. Where it helps, the learned subspace carries the gain: once the subspace is
learned, a probe on it is as robust as simulating interventions and costs no
more than any probe. IIA does not reliably identify subspaces that are useful for correctness prediction. On gpt-oss, well-converged subspaces were no better than random ones, so the
subspace has to be validated on held-out data.

\section{Threats to Validity}
\label{sec:threats}
\paragraph{Construct validity}
Correctness is defined by the benchmarks' tests, which can miss errors. The
judging prompt asks for a single token without reasoning; gpt-oss, a
reasoning model, judges moderately  better with reasoning: on 498 program-target
candidates of the language sets, medium-effort reasoning (median 1{,}101
tokens) raises its balanced accuracy from 62.4\% to 67.8\% and the AUROC of its
\texttt{Yes} probability from 0.68 to 0.73, still below the trained internal
methods (Table~\ref{tab:program-summary}), at the cost of the reasoning
tokens. 

\paragraph{Internal validity}
The trained methods choose among different numbers of layers: the probe and
LAT among all 25--49, the causal methods among the 4--7 layers at which DAS
was trained, since each subspace takes several GPU hours. 
A wider choice can find a better layer, but under shift the layer best in distribution need not transfer, so the effect can go either way. Under shift, however, the causal checkpoints often lie above the probe even at the probe's best layer (Figure~\ref{fig:verdict-profile}).
Counterfactual simulation uses one fixed set of 16 bases;
drawing them from different problems changed the AUROC of one checkpoint by at
most 0.01 AUROC. 
Reruns of the same configuration on different machines differ by up to 0.02 AUROC on identical rows, so we do not interpret smaller differences between single cells.
EigenScore needs five non-empty generations per problem (ten on BCB), so the problem-level
comparison leaves out the problems with fewer: 40 for Qwen3-Coder, none of them
solved, and 143 for gpt-oss, 47 of them solved.

\paragraph{External validity}
We study two open-weight models on competitive-programming problems from LiveCodeBench, on a subset of the same problems in Python, C++ and Java, which
isolates the language, and on more practical Python tasks from BigCodeBench.
Other models, sampling settings or repository-level tasks may behave differently.

\section{Conclusion}
\label{sec:conclusion}
We compared execution-free predictors of code correctness from four families
under benchmark and language shift, for two targets: whether a program is
correct and whether the model's verdict on it can be trusted. Causal
localization, adapted from natural-language question answering, gives the
most robust predictor of the model's verdict under language shift on one of
two models, and its gain lies in the learned subspace, so a simple probe on it
suffices. 
Beyond that, how well a predictor generalizes depends on the target
and the type of shift: no verdict predictor transfers reliably between
benchmarks, while program correctness transfers better. Much of what the predictors
recognize is the problem rather than the candidate, which makes them better suited to deciding which problems to verify than to choosing among candidates. 
Future work should test whether these signals can gate test execution and stronger judges in agentic settings, and whether a few labels from the target setting suffice to adapt them. 


\bibliographystyle{plain}
\bibliography{ood_correctness}
\end{document}